# Compaction and flow rule of oxide nanopowders

G.Sh. Boltachev[a,*], K.E. Lukyashin[a], A.L. Maximenko[b], R.N. Maksimov[a,c], V.A. Shitov[a], M.B. Shtern[b]

[a] *Institute of Electrophysics, Ural Branch of RAS, Amundsen St. 106, Ekaterinburg 620016, Russia*

[b] *Frantsevich Institute for Problems of Materials Science, NAS Ukraine, Krzhizhanovsky St. 3, Kyiv 03680, Ukraine*

[c] *Ural Federal University named after the first President of Russia B.N. Yeltsin, Mira St. 19, Ekaterinburg 620002, Russia*

[*] *Corresponding author, tel.: +7 343 267 87 76, e-mail: grey@iep.uran.ru*

**Abstract**

Transparent $Al_2O_3$ ceramics have attracted considerable interest for use in a wide range of optical, electronic and structural applications. The fabrication of these ceramics using powder metallurgy processes requires the development of theoretical approaches to the compaction of nanopowders. In this work, we investigate the compaction processes of two model granular systems imitating $Al_2O_3$ nanosized powders. System I is a loosely aggregated powder, and system II is a powder strongly inclined to agglomeration (for instance, calcined powder). The processes of isostatical (uniform), biaxial, and uniaxial compaction as well as uniaxial compaction with simultaneous shear deformation are studied. The energy parameters of compaction such as the energy change of elastic interparticle interactions and dispersion interactions, dissipative energy losses related to the processes of interparticle friction, and the total work of compaction are calculated for all the processes. The nonapplicability of the associated flow rule to the description of deformation processes of oxide nanopowders is shown and an alternative plastic flow rule is suggested. A complete system of determining the relationship of the flow including analytical approximations of yield surfaces is obtained.

## 1. Introduction

In recent years, polycrystalline ceramic materials based on refractory oxides such as $Al_2O_3$ [1-8] and $Y_2O_3$ [7,9,10] have enjoyed considerable research interest as potential candidates for various optical applications. In particular, high heat conductivity is a known advantage of aluminium oxide as a laser medium [6]. The average grain size and the sizes of residual pores must be decreased by up to the values of about 10 nm in order to achieve high optical quality and a desirable mechanical resistance of alumina ceramics [2,6]. Hence, the development of nanotechnologies and, in particular, the fabrication of nanostructured ceramics using powder metallurgy processes is closely connected with a growing interest in transparent ceramics [4,6-10]. The cold compaction of nanosized powders is the most commonly used processing step for powder metallurgy [7-11]. In contrast to micron (or larger) powders, nanopowders possess a number of unexpected properties [11-13], which have an influence on powder compaction and subsequent sintering. Primarily, they have a pronounced size-related effect: the smaller the particle size, the harder it is to compact the powder [12,13]. In some cases the pressure of several GPa is required to obtain a desired density of oxide nanopowder during the cold compaction process [9,11-13]. In addition, as demonstrated in [12,13], the nanopowders of oxide materials are beyond the associated flow rule and are weakly sensitive to the compaction method since the difference in the density after isostatical (uniform) and uniaxial pressing does not exceed 1%.

The rapid development of experimental methods and further success in the fabrication of nanostructured oxide ceramics require the corresponding development of theoretical conceptions of the mechanical properties of the nanopowder compact. In the space of stress tensor invariants the yield surface has a convex form of an elliptical type [12,13] that is supposed to rely on the theory of plastically hardening porous bodies as a continuum approach for describing the properties of the nanopowder [11,14]. Thus, obviously, a number of conceptions and a terminology related to the theory attain a conditional character, in particular, that the plasticity of a powder body is correlated with the processes of the mutual sliding and rearrangement rather than the deformation of the individual particles. The features of the nanopowder body demand a serious inspection of the main conceptions of the theory and verification of its results towards the properties of the described body. A full-scale experiment is unable to give comprehensive information on the characteristics of the powder system and the evolution of its properties during the compaction processes. In fact, from the experiment we only find the powder compaction curve under well-defined pressing conditions.

Far more detailed information can be obtained within the framework of the microscopic investigation presented in this paper, i.e. a computer simulation of the powder by the granular dynamics method [12,13]. For research objects, we use two monosized model systems (the particle diameter $d$ = 10 nm) corresponding to alumina nanopowders, which have a weak (system I) and strong (system II) inclination to agglomeration [12]. Such powders are produced by the Institute of Electrophysics (Ekaterinburg, Russia) by electric explosion of wires [15] and by laser ablation [9,16]. The individual particles have a spherical form and high strength properties. The particle sphericity, high strength, and nonsusceptibility to plastic deformation make granular dynamics a promising and adequate tool of theoretical analysis.

## 2. Calculation

In this study, the processes of cold quasistatic pressing such as isostatical compaction (process $A$), biaxial ($B$) and uniaxial ($C$) compaction, and uniaxial compaction with simultaneous shear deformation ($D$) are simulated by the granular dynamics method in 3D geometry. These processes are performed by simultaneous changes of selected sizes of a model cell (all sizes when simulating the process $A$; the cell height $z_{cell}$ when simulating the uniaxial compaction; and so on) and proportional rescaling of the appropriate coordinates of all the particles. After each step of deformation, the new equilibrium locations of the particles are determined during a large number $N_{iter}$ of equilibration steps (several hundred, as a rule). This work is a continuation and development of studies presented in Refs. [12,13], in which the above-mentioned processes and all the interparticle interactions used in the model are described in detail. Here we point out that the interactions of particles include elastic repulsion, tangential "friction" forces, resistance to relative rotation of particles and their "rolling", dispersion forces of attraction, and the possible formation of strong bonds of a chemical nature for system II. The parameters of particles material correspond to α phase of aluminium oxide. Examples of this are Young's modulus $E$ = 382 GPa, Poisson's ratio $v$ = 0.25, and the dispersive energy of intermolecular attraction $\varepsilon$ = 1224 $k_B$ ($k_B$ is the Boltzmann's constant). The number of particles in a model cells $N_p$ = 1000, and the maximum compaction pressure $p_{max}$ = 5 GPa. Such high pressure is not achievable using the experimental equipment for static compaction, but could be obtained by inertial effects during the processes of magnetic pulsed compaction [11].

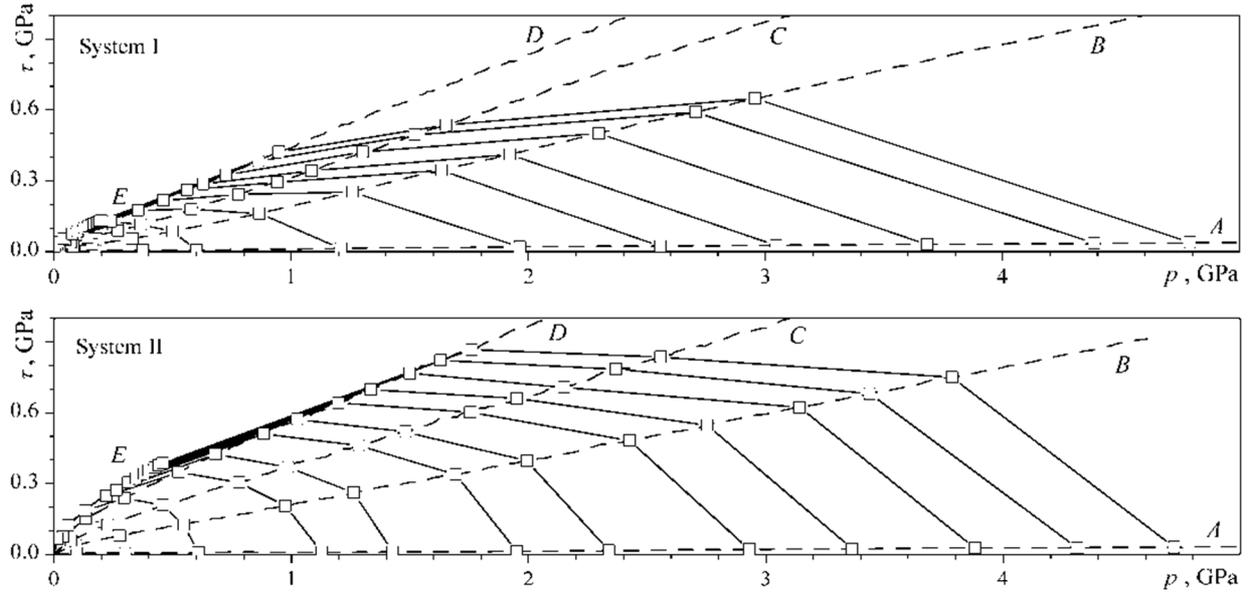

Fig. 1. Levels of the yield surface in coordinates of "hydrostatic pressure – deviator intensity" of the stress tensor for system I (from the top) and system II (from below). Solid lines correspond to the values of the unloading density $\rho_u$ = 0.4, 0.5, 0.55, 0.58, 0.59, 0.60, 0.605 (both systems); 0.608, 0.61, 0.612, 0.614 and 0.615 (system I); 0.61, 0.613, 0.616, 0.618, 0.62 (system II). Dashed lines correspond to the processes of isostatical (*A*), biaxial (*B*), uniaxial (*C*) compactions, uniaxial compaction with shear deformation (*D*), and pure shear (*E*).

In contrast to earlier studies [12,13], in this work we have added minor changes to the numerical algorithm in order to increase accuracy during the calculation of the compaction curves. The number of independent computer experiments for statistical averaging when plotting the calculated curves was increased by four times (from 10 to 40); the maximum number of equilibration steps $N_{iter,max}$ for achieving new equilibrium locations of the particles was increased by two times (up to 2000); and the minimum value of the particle attractive force $f_{a,min}$ (where the cutting of the dispersive potential occurred) was decreased from $5\times10^{-6}$ to $1\times10^{-6}$ (in reduced units: $f_{red} = f/(Ed^2)$). These minor modifications of the numerical algorithm led to small shifts in the compaction curves presented in [13]. The shifts are not visually recognizable (the density shifts are about 0.2%), but they have sufficient amplitude in the pressures (about 100 MPa at the maximum compaction pressures). As a result, this led to a significant change in the yield surfaces of the investigated systems (Fig. 1) corresponding to the defined values of compact unloading density $\rho_u$, i.e., the density of compact after the removal of the external load. Primarily, this is related to System II (with strong bonds). In particular, process *D* now has higher values of the deviator intensity $\tau$ compared with process *C* throughout the whole range of investigated loads up to pressures of $p_{max}$ = 5 GPa. It was previously observed only in a loading range below 3 GPa [13]. Note that the process of pure shear *E* as shown in Fig. 1 was not calculated in this work and the corresponding points were taken from Ref. [13]. The analysis performed makes it clear that the primary factor resulting in the change of yield surfaces is the increase of the $N_{iter,max}$ parameter. This fact indicates that the yield surface is a very sensitive property of the simulated systems and a reliable calculation of the yield surface requires a high level of accuracy in the achievement of an equilibrium state for all the particles after every deformation step of a model cell.

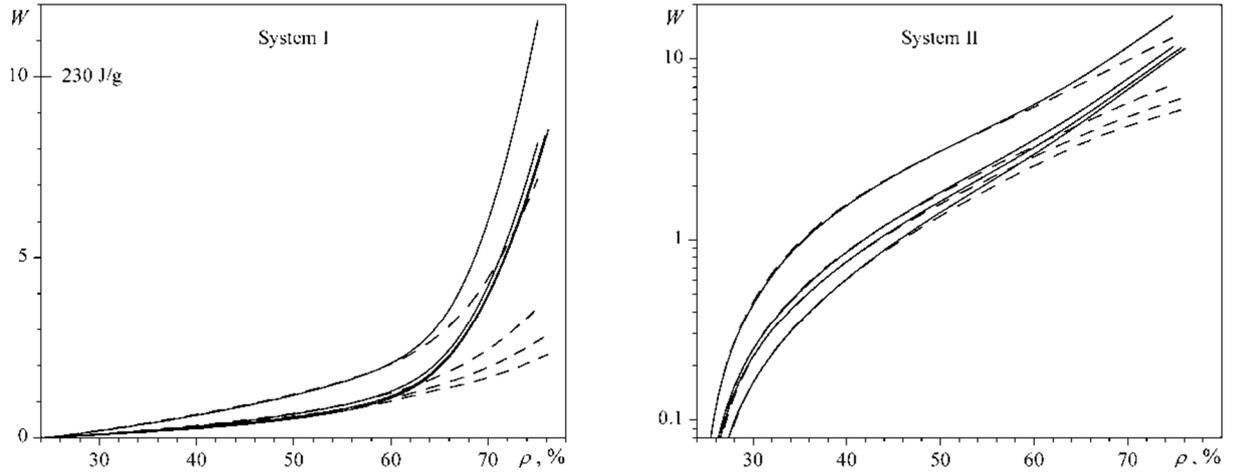

Fig. 2. Compaction work $W_{out}$ (in reduced units; $1 \cong 23$ J/g) for the model systems I (on the left) and II (on the right) at processes A–D (see Fig. 1) (solid lines bottom-up). Dashed lines are the dissipative energy losses $W_f$ due to interparticle friction.

## 3. Results and discussion

*3.1. Energy parameters of the compaction processes*

Despite the change of the yield surfaces noted in Sec. 2, one of the main conclusions of Ref. [13], i.e. nonapplicability of the associated flow rule to the oxide nanopowders, remains valid. To find another flow rule of the compacted medium, which could be used instead of the associated rule, energy parameters such as the energy of the particles' elastic strains $E_{el}$, the energy of dispersive attractions $E_a$, and dissipative energy losses related to the processes of interparticle friction $W_f$ were investigated in this study. Fig. 2 shows the total work of powder compaction $W_{out} = \Delta E_{el} + \Delta E_a + W_f$ and its part coming from dissipative losses to the interparticle friction $W_f$. As can be seen, up to densities of about 60% the compression work is mainly expended to overcome the interparticle friction. Here the significant rearrangement of particles occurs, the average coordination number increases notably, and the effective compaction of a powder body is performed. A further increase in pressure leads to a considerable growth of the elastic stresses, i.e. a powder structure is "stuck" in a fixed configuration. Subsequent deformation of a model cell (at $\rho > 0.6$) becomes more and more elasto-reversible, and the unloading density of the compact $\rho_u$ practically stops increasing. In addition, Fig. 2 shows that the values of $W_{out}$ and $W_f$ significantly depend on the process being performed. This indicates that the powder could not be characterized by a universal relationship $W(\rho)$ or $W(\rho_u)$ and, as a consequence, it is impossible to expect a coincidence in the space of stress tensor invariants with the yield surfaces satisfying the $\rho_u$ = const condition and the levels of the plastic potential ($W_f$ = const) determining the energy dissipation rate.

Fig. 3 presents the levels corresponding to the fixed values of the total strain $W_{out}$ and the dissipative losses $W_f$. On a plane $(p,\tau)$, where $p$ is the hydrostatic pressure and $\tau$ is the deviator intensity of the stress tensor, these levels are well-approximated by the following equation:

$$\tau(p) = (c_1 + c_2 p)\sqrt{(1-\alpha)^2 p_A^2 - (p - \alpha p_A)^2}, \qquad (1)$$

where the hydrostatic pressure at the isostatic compaction $p_A$ is used as a parameter characterizing the fixed level, parameter $c_2 > 0$ for levels of $W_{out}$ and $c_2 = 0$ for levels of $W_f$. An analytical description of $W_{out}$ and $W_f$ levels by equation (1) simplifies their theoretical analysis. To the left of the maximum (where $\partial W_{out}/\partial p = 0$, or $\partial W_f/\partial p = 0$) the levels for $W_{out}$ as well as for $W_f$ form a common envelope curve, which nearly coincides with the curve of the process $D$. In our opinion, it is necessary to identify this envelope as a limiting curve corresponding to a known failure surface [17]. Therefore, the level lines in Fig. 3 do not continue up to intersect with the $p$-axis at the value of $p = (2\alpha-1)/p_A$ but come to an end when reaching the envelope.

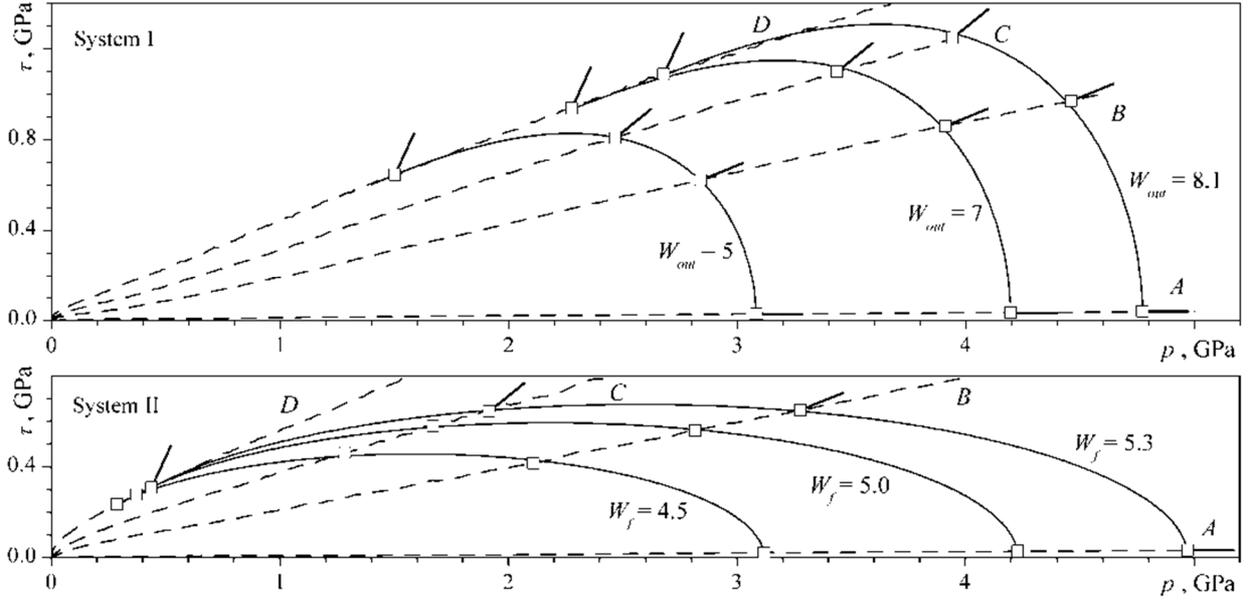

Fig. 3. Levels of total work $W_{out}$ for the compaction of system I (from the top) and dissipative energy losses $W_f$ in system II (from below) in coordinates of "hydrostatic pressure – deviator intensity" of the stress tensor. Dots are the simulation results, lines are approximating curves by Eq. (1). Dashed lines correspond to $A$–$D$ processes (see Fig. 1). Arrows show the "direction" of deformation, i.e. of $(e,\gamma)$ vectors.

In addition, Fig. 3 shows the directions of vectors $(e,\gamma)$, where $e$ is the first invariant and $\gamma$ is the deviator intensity of the strain rate tensor. As a "rate" here we have implied the derivation with respect to some parameter, which determines the state of the system in the compaction process, rather than time differentiation (since all the simulated processes are quasistatic). In the framework of conventional plasticity theories the "$W_f$ = const" levels are identified with the iso-surfaces of plastic potential and the vectors $(e,\gamma)$ should be normal to these surfaces in accordance with the generally accepted associated rule. However, Fig. 3 reveals that the associated rule is not valid regarding both the levels of $W_f$ = const and the levels of the total work $W_{out}$ = const. It was demonstrated earlier in Ref. [13] that this rule is also inapplicable regarding the yield surfaces ($\rho_u$ = const) presented in Fig. 1.

*3.2. Flow rule of the oxide nanopowders*

Analysis of the deviations of the deformation directions from the directions prescribed by the associated rule, i.e. the deviations of $(e,\gamma)$ vectors from the normal directions to the surfaces of $W_f$ = const (Fig. 3) or to the yield function surfaces (Fig. 1), permits us to suggest an alternative flow criterion of oxide nanopowders as follows:

$$(e,\gamma) = (1-\omega)\lambda_1' \nabla\Phi + \omega \lambda_2(p',\tau') \ , \qquad (2)$$

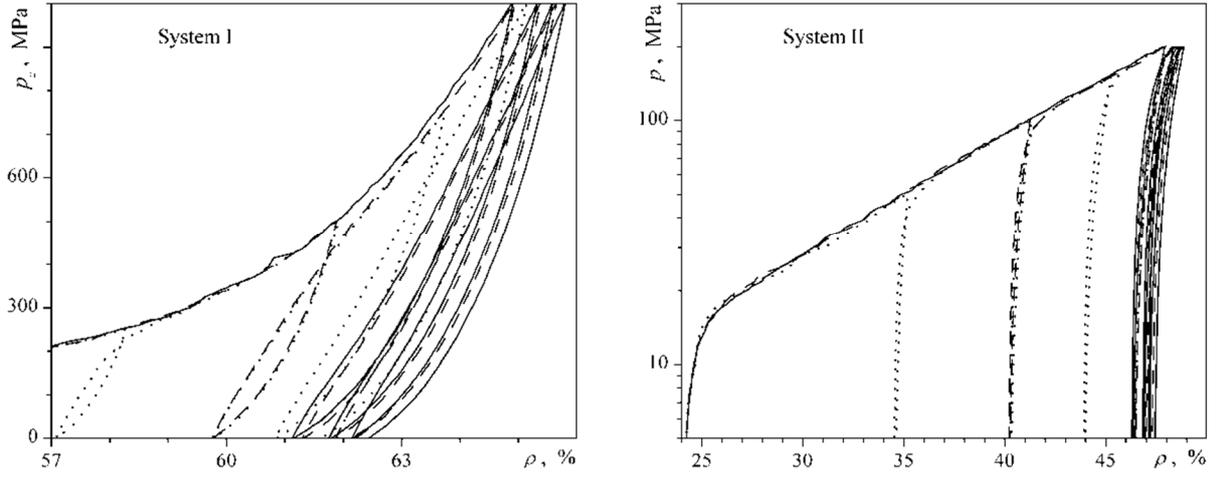

Fig. 4. Dependences of pressure on density for system I (on the left; uniaxial compaction, $p_{max}$ = 1 GPa) and for system II (on the right; isostatical compaction, $p_{max}$ = 200 MPa) at the multiple loading. Solid lines correspond to four-time loading up to the maximum pressure $p_{max}$; dashed lines correspond to the first loading up to 0.5 $p_{max}$ and three-time loading up to $p_{max}$; and dotted lines correspond to the subsequent loadings up to 0.25 $p_{max}$, 0.5 $p_{max}$, 0.75 $p_{max}$, and $p_{max}$.

where $\omega$ is the weight coefficient determining the influence of the process being performed on the "direction" of the deformation occurring in the system, a dash indicates the differentiation with respect to a parameter characterizing the change of the system state during the compaction process, and $\lambda_2$ is the dimension factor given by the equation:

$$\lambda_2 = \lambda_1' \frac{|\nabla \Phi|}{|(p', \tau')|} = \lambda_1' \sqrt{\left[\left(\frac{\partial \Phi}{\partial p}\right)_\tau + \left(\frac{\partial \Phi}{\partial \tau}\right)_p\right] \Big/ \left[p'^2 + \tau'^2\right]}. \qquad (3)$$

Any function concerned either with levels of $W_f$ = const or with the yield function can be used as $\Phi$ potential in Eqs (2) or (3). In our further analysis, we prefer the yield function since it represents a more obvious, precise, and reliable characteristic of a powder body.

The latter is confirmed by an additional investigation into the multiple loading of the model systems I and II. We simulated a four-time loading of both systems using three various schemes. In the first scheme, the model systems were loaded multiple times up to the maximum pressure $p_{max}$ and then unloaded down to a full pressure release. In the second scheme, the first loading was performed up to half of the maximum pressure ($p_{max}$ / 2), and the further three loadings up to $p_{max}$. In the third scheme, the loadings followed by full pressure releases were performed up to 0.25 $p_{max}$, 0.50 $p_{max}$, 0.75 $p_{max}$, and $p_{max}$. The processes of uniaxial and uniform (isostatical) compaction at the maximum pressures $p_{max}$ = 200 MPa and 1 GPa were investigated. Fig. 4 presents the results of these numerical experiments. In particular, it can be seen that the final density of the compact does not increase by conducting the intermediate loadings. Conversely, the intermediate loadings up to lower values of the compaction pressures lead to a lower final density of the compact than multiple loadings up to the maximum pressure. However, the difference in density does not exceed 1%. In general, the compaction curve practically does not depend on the intermediate "loading-unloading" stages, and the $p(\rho)$ dependence quickly returns to a certain well-determined curve at the further loading. We can therefore conclude that regardless of the intermediate "loading-unloading" stages and, in particular, regardless of the initial state of the powder, the yield function is a reliable characteristic of the investigated powder material.

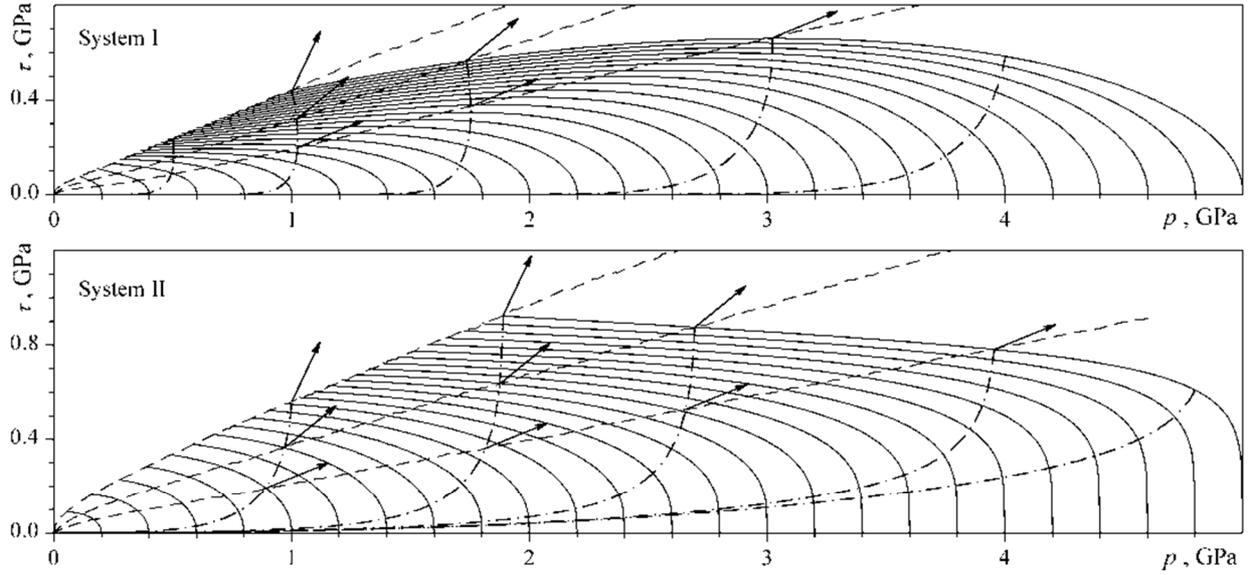

Fig. 5. Levels of yield surface plotted by Eqs. (4) and (5). Dashed lines correspond to the processes $B–D$ (Fig. 1). Dash-and-dot lines show the direction of the yield function gradient ("lines of flow"). Arrows show the "direction" of the deformation, i.e. of $(e,\gamma)$ vectors.

Employing the yield function as the $\Phi$ parameter included in the flow rule of a powder body (2) requires a convenient analytical description of the levels ($\rho_u$ = const) as presented in Fig. 1. For system I, a satisfactory description of the calculation data is achieved by the following equations:

$$\tau(p) = p_A \left[ c_1 + c_2 \left(\frac{p}{p_A}\right) + c_3 \left(\frac{p}{p_A}\right)^2 \right] \sqrt{1 - \frac{p}{p_A}}, \quad c_i = c_{i0} + \frac{c_{i1}}{1 + p_A/c_{i2}} + \frac{c_{i3}}{(1 + p_A/c_{i4})^2} + \frac{c_{i5}}{(1 + p_A/c_{i6})^3}, \quad (4)$$

and for system II the approximate equation takes the form:

$$\tau(p) = p_A \left[ c_1 + c_2 \left(\frac{p}{p_A}\right) \right] \sqrt{1 - \frac{p}{p_A}} + p_A c_3 \left(1 - \frac{p}{p_A}\right)^{1/4}, \quad c_i = c_{i0} + \frac{p_A}{c_{i1}} + \left(\frac{p_A}{c_{i2}}\right)^2 + \left(\frac{p_A}{c_{i3}}\right)^3. \quad (5)$$

As a result of regression analysis, the following parameter values were obtained (absent coefficients are equal to zero, all the dimension coefficients are given in GPa): $c_{13} = 7.69$, $c_{15} = -5.07$, $c_{14} = 0.57$, $c_{16} = 1.24$, $c_{20} = 1.243$, $c_{21} = -44.744$, $c_{22} = 0.209$, $c_{23} = 99.382$, $c_{24} = 0.580$, $c_{25} = -62.990$, $c_{26} = 0.908$, $c_{30} = -2.707$, $c_{31} = 111.989$, $c_{32} = 0.296$, $c_{33} = -270.087$, $c_{34} = 0.784$, $c_{35} = 168.357$, and $c_{36} = 1.266$, for Eq. (4); $c_{10} = 0.456$, $c_{11} = -5.905$, $c_{12} = 10.365$, $c_{20} = 0.364$, $c_{21} = -5.814$, $c_{22} = 4.968$, $c_{23} = -6.037$, $c_{31} = 21.54$, and $c_{32} = 13.70$ for Eq. (5).

Fig. 5 shows the levels of the yield functions for systems I and II plotted using Eqs. (4), (5) and the parameter values $p_A$ from 0 to 5 GPa. In contrast to the energy levels (see Fig. 3), the limiting fracture surface (conventionally matching with the curve of the process $D$) is not the envelope for the levels of the yield function. However, an extension of the yield surface over the fracture surface is meaningless. Fig. 5 also presents the "line of flow" (dash-and-dot lines) where the direction is aligned with the direction of the yield function gradient $\nabla \Phi = (\partial \Phi / \partial p, \partial \Phi / \partial \tau)$. According to the traditional associated rule, these lines should determine the direction of deformation in a system, i.e. $(e,\gamma) = \lambda_1' \nabla \Phi$. Owing to the use of analytical expressions (4) and (5) the non-orthogonality of $(e,\gamma)$ vectors to the levels of the yield function can be distinctly seen in Fig.

5. Vectors (*e*,*γ*) deviate significantly from the directions prescribed by the associated rule to the directions of curves (dashed lines) corresponding to the compaction processes being performed.

Proceeding from the expressions for the invariants (2) to the deformation and stress tensors, a general form of the flow rule of the oxide nanopowders can be written as follows:

$$\varepsilon'_{\alpha\beta} = (1-\omega)\,\lambda'_1 \cdot \frac{\partial \Phi}{\partial p^{\alpha\beta}} + \omega\,\lambda_2 \left( p'\frac{\partial p}{\partial p^{\alpha\beta}} + \tau'\frac{\partial \tau}{\partial p^{\alpha\beta}} \right), \qquad (6)$$

where

$$\frac{\partial \Phi}{\partial p^{\alpha\beta}} = \left(\frac{\partial \Phi}{\partial p}\right)_{\tau}\frac{\partial p}{\partial p^{\alpha\beta}} + \left(\frac{\partial \Phi}{\partial \tau}\right)_{p}\frac{\partial \tau}{\partial p^{\alpha\beta}}, \qquad \frac{\partial p}{\partial p^{\alpha\beta}} = \frac{\delta_{\alpha\beta}}{3}, \qquad \frac{\partial \tau}{\partial p^{\alpha\beta}} = \frac{1}{\tau}\left(p_{\alpha\beta} - p\cdot\delta_{\alpha\beta}\right),$$

and $\delta_{\alpha\beta}$ is Kronecker's delta. The first summand from the right in Eq. (6) corresponds to the associated rule, and the second summand determines the influence of the process being performed. Fig. 5 shows that the weight coefficient of this influence $\omega$ is not constant. It can be suggested that $\omega$ is a monotonically decreasing function of the ratio $\tau'/p'$. The analysis of numerical data presented in Fig. 5 allows us to suggest the approximation of this function as follows:

$$\omega = 1 - \omega_2 \left(\frac{\tau'}{p'}\right)^2, \qquad (7)$$

where $\omega_2 = 3$ for system I and $\omega_2 = 4$ for system II. The approximation (7) closes the system of previous equations, which determines the change of strain tensor of the powder system at a prescribed external load, i.e. at a prescribed increment of the stress tensor components.

## 4. Conclusions

The energy costs (a work of compaction, dissipative energy losses, and so on) during the compaction in different conditions, such as uniform, biaxial, and uniaxial compactions, and uniaxial compaction with simultaneous shear deformation, have been studied for the model systems corresponding to nanosized alumina powders with a weak (system I) and strong inclination to agglomeration (system II). It has been found that the compaction work depends significantly on the process conditions and is not a single-valued function of the powder density. It is evident that the nanopowders require independent characterization not only by a set of yield surfaces but also by using a set of energy levels. A known associated flow rule was found to be inapplicable to the nanosized oxide powders. Another rule, which allows the prediction of the character of the deformation processes in a system, was suggested in place of the associated rule. According to the suggested rule, the strain rate tensor is determined not only by the direction of the gradient vector of the yield function (the associated flow rule), but also by the direction of the "vector" determining the changes of the stress tensor components in the compaction process being performed. The ratio of the contributions from these two vectors is assigned by the weight coefficient $\omega$. A complete system of equations, which determines the changes of the strain tensor components in the powder systems at the prescribed external loading, was obtained. In addition to the flow rule, this system contains an approximation of the yield surface levels and a dependence of the weight coefficient $\omega$ on the stress tensor invariants.

It should be noted that four investigative processes may not be enough. A more reliable construction of yield surface levels and, especially, a determination of the relationship between the weight coefficient $\omega$ and the stress tensor invariants require additional investigation. In particular, the $\omega$ coefficient can be a function of a larger set of parameters, such as $\omega = \omega(\tau'/p'; p, \tau)$. The possibility of generalizing the obtained ratios for other powder systems, with the characteristics of the individual particles and interparticle interactions differing from the alumina particles, remains unclear. These problems are the scope of further research.

**Acknowledgements**

The work has been fulfilled in the frame of state task project No. 0389-2014-0002 and supported by RFBR (project Nos. 14-08-90404-Ukr_a and 16-08-00277), National Academy of Science of Ukraine (project No. 25-08-14), and Act 211 Government of the Russian Federation (contract No. 02.A03.21.0006).